\documentclass[12pt]{iopart}
\usepackage{epsfig}
\begin{document}

\newcommand{\non}{\nonumber}
\newcommand{\be}{\begin{equation}}
\newcommand{\ee}{\end{equation}}
\newcommand{\bq}{\begin{eqnarray}}
\newcommand{\eq}{\end{eqnarray}}

\title
[From the superfluid to the Mott regime and back]
{From the superfluid to the Mott regime and back: triggering a non-trivial 
dynamics in an array of coupled condensates
}

\author{P Buonsante\dag\,,
R Franzosi\ddag\,, and V Penna\dag
\footnote[3]{To
whom correspondence should be addressed (franzosi@df.unipi.it)}
}

\address{\dag\ Dipartimento di Fisica, Politecnico di Torino,
and INFM, UdR Torino, \\
C.$^{so}$ Duca degli Abruzzi 24,
I-10129 Torino, Italy}

\address{\ddag\
Dipartimento di Fisica dell' Universit\`a di Pisa,
INFN, Sezione di Pisa,\\
 and INFM UdR di Pisa \\
Via Buonarroti 2, I-56127 Pisa, Italy}

\begin{abstract}
We consider a system formed by an array of Bose-Einstein 
condensates trapped in a harmonic potential with a superimposed 
periodic optical potential. 
Starting from the boson field Hamiltonian, appropriate to describe dilute gas 
of bosonic atoms, we reformulate the system dynamics within the 
Bose-Hubbard model picture. Then we analyse the effective dynamics of the 
system when the optical potential depth is suddenly varied according to a 
procedure applied in many of the recent experiments on 
superfluid-Mott transition in Bose-Einstein condensates.

Initially the condensates' array generated in a weak optical 
potential is assumed to be in the superfluid ground-state
which is well described in terms of coherent states.  
At a given time, the optical potential depth is suddenly increased
and, after a waiting time, it is quickly decreased so that the initial 
depth is restored.
We compute the system-state evolution and show that the
potential jump brings on an excitation of the system,
incorporated in the final condensate wave functions, whose
effects are analysed in terms of two-site correlation functions 
and of on-site population oscillations.
Also we show how a too long waiting time can destroy completely
the coherence of the final state making it unobservable.
\end{abstract}


\maketitle

\section{Introduction}
Bose-Einstein condensation, originally observed in a
dilute atomic gas trapped in a harmonic potential \cite{A:MHAnderson,A:Brandley},
is today obtained in a variety of experimental configurations.
Experimental efforts have allowed to realize setup
in which condensates are achieved into one-,
two- or three-dimensional optical lattices \cite{A:BAnderson,A:Marzlin,A:Morsch,A:Morsch2},
that is arrays of microscopic potentials
induced by ac Stark effect due to interfering laser
beams. Such microscopic potentials are often superimposed to
the trapping harmonic one and give rise to a fragmentation
of the condensate.

In the very recent experiments on Bose-Einstein condensates
(BECs) in optical lattices, dynamically active states have been
generated either by accelerating or by tilting the optical lattice
or by shifting the harmonic potential trap \cite{A:BAnderson,A:Marzlin,A:Morsch,A:Morsch2}.
Nevertheless, in such experimental realizations essentially 
classical/superfluid regimes have been explored and the corresponding 
dynamics results to be quite well described by the discrete Gross-Pitaevskii
equation (GPE). Phenomena as Bloch oscillation, nonlinear Landau-Zener
tunneling, Josephson junction current, can be explained in terms of
the band structure entailed by the GPE in the spirit of 
solid-state physics \cite{A:Berg}.

The opposite regime, where a low number of bosons per well or a strong
optical potential require a quantum description of the dynamics,
has been recently explored in some experiments \cite{A:GreinSC,A:Orzel}.
A strong  interplay between quantum and classical regime takes place
in such experiments when, for example, the quantum phase transition 
from the superfluid to the Mott insulator regime is generated, or when 
the collapse and revival of the bosonic wave functions is observed.

In the present paper we consider an experimentally realistic system
constituted by a dilute gas of $N$ ultracold bosonic atoms, trapped 
in a harmonic potential and loaded into an one-dimensional optical 
lattice of $M$ wells. 
At the beginning, we derive an effective dynamics by reformulating the 
second-quantized many-body Hamiltonian, that well describes the 
dynamics of this system, within a generalised Bose-Hubbard model (BHM) 
picture. In such a way we know the effective
Hamiltonian dynamical parameters as a function of the microscopic 
system constants as well as of external variables such as the 
magnetic-optical potential strength. Based on such effective picture, 
we study the system dynamics when the optical potential depth is 
quickly varied.

Initially, we consider the situation where the system 
is in the ground-state involved by a weak optical potential.  
Clearly, the discrete GPE should be applied in this regime to 
recognize the ground-state configuration \cite{A:Polkov,A:Celeghini,
A:Chernyak}. At time 
$t=0$ the lattice potential depth is suddenly increased, so that 
the tunneling amplitude between neighbouring wells quickly drops
to zero. Consequently, the system enters in the {\em Mott regime} 
in which the time evolution requires a quantum description.
After a waiting time $t^\prime$, the optical potential depth is
suddenly decreased to the initial value. From this time on, the 
discrete GPE gives a satisfactory description of the system time 
evolution.
The initial conditions for the following mean-field
evolution driven by the GPE thus stem from the quantum state 
emerging from the Mott regime.
We compute the quantum state describing the system as a function
of time and of the Hamiltonian parameters, and show that 
the phase shift between 
the condensates of neighbouring wells exhibits a strong time dependence
while the well populations undergo oscillations. The system is thus taken 
in an excited state \cite{A:Band,A:TrombL2,A:Menotti2}.  
Further, we show both that the final state coherence dramatically depends
on the waiting time $t^\prime$, and that the site wave-function
phase coherence is destroyed when $t^\prime$ is increased.

\section{Space-mode approximation.}
The Hamiltonian operator for a dilute gas of bosonic atoms in a 
harmonic trapping potential $V_{H} ({\bf r})=\Sigma^3_{j=1} m 
\Omega^2_j r^2_j /2$ 
with the additional one-dimensional optical 
lattice potential 
$V_{L} ({\bf r})=  \hbar^2 \omega^2 \sin^2(k r_1)/(4 E_r)$, 
($k$ is the laser mode and $E_r =\hbar^2 k^2/(2 m)$
is the recoil energy) has the following form
$$
{\hat {H}} = \! \int \! d^3{\bf r} 
\hat\psi^+({\bf r}) 
\left [ -\frac{\hbar^2}{2m}\nabla^2 
+ V_{H} ({\bf r})+ V_{L} ({\bf r}) \right ] \hat\psi({\bf r}) + 
\frac{4\pi \hbar^2 a_s}{2m} \!
\int \! d^3{\bf r} 
\hat\psi^+({\bf r})\hat\psi^+({\bf r})
\hat\psi({\bf r})\hat\psi({\bf r}),
$$
where $\hat\psi ({\bf r})$ ($\hat\psi^{+}({\bf r})$) 
is the annihilation (creation) boson-field 
operator for the atoms in a given internal state, $a_s$ is the $s$-wave 
scattering length and $m$ is the atomic mass.
The space-mode approximation~\cite{A:Milburn}, 
which allows us to reformulate the system
dynamics within the BHM picture, is performed as follows.
Let $V_j$ be the parabolic approximation to $V = V_L + \Sigma^3_{j=2} m 
\Omega^2_j r^2_j /2$ in ${\bf r}_j= (j \pi/k,0,0) $
the locations of $V$ local minima. 
We assume the energies involved in the system dynamics to be small
compared to the excitations of the single well ground-state.
Thus we can expand the boson field operators
in terms of Wannier functions $\hat\psi({\bf r},t) = \sum_{{ j}} 
{u}^*_{ j}({\bf r}) \hat a_{ j}(t)$. 
In the last equation ${ j}$ runs on the optical lattice sites,  
${u}_{ j}$ is the single-particle ground-state mode 
of $V_{ j}$ with energy eigenvalue  $\epsilon (\omega) = \hbar (\omega 
+ \Omega_2 + \Omega_3)/2$.
By substituting the previous expression of $\hat\psi({\bf r},t)$ in the
previous Hamiltonian 
and keeping the lowest order in the overlap 
between the single-well modes, we find the Bose-Hubbard Hamiltonian
\be
H=  \sum_{ i}[\, U n_{ i}(n_{ i}-1) + \lambda_{ i} n_{ i} ] -
{\frac{T}{2}} \sum_{<ij>}  \,
\left (a^{+}_{ i} a_{ j} + h.c. \right ) \, ,
\label{BHH}
\ee
where the operators $n_{ i}=a^{+}_{ i} a_{ i}$ count the number
of bosons at the $i$-site of the lattice and the annihilation and creation
operators $a_{ j}$ and $a^{+}_{ j}$ satisfy the standard commutation
relations $[ a_{i}$, $a^{+}_{j}] = \delta_{i,j}$.
In Hamiltonian \ref{BHH} parameters are defined as follows.
$U:= a_s \Omega_0 \sqrt{m \hbar \Omega_0 /(2 \pi)}$ is the
strength of the on-site repulsion, in which we have set $\Omega_0
= \sqrt[3]{\omega \Omega_2 \Omega_3}$. 
The site external potential is $\lambda_{ j}:=\epsilon (\omega) 
+{ j}^2 \pi^2 \hbar^2 \Omega^2_1/(4 E_r)$, 
and $T:=-2 \int d^3{\bf r } \bar{u}_{ j} [V - 
V_{ j\pm { 1}}]u_{{ j} \pm { 1}}$ 
is the tunneling amplitude between neighbouring sites.
The indices $i,j \in Z$ label the local minima 
$x_j = \pi j/k$ of $V$ throughout the lattice and 
$V_{j} = m \omega^2 (r_1-x_j)^2/2 
+ \Sigma^3_{\ell=2} m \Omega^2_\ell r^2_\ell /2$. The total number of 
bosons $N=\Sigma_j n_j$ is conserved.
>From now on we shall consider a gas of repulsive atoms, thus $U >0$.

BHM (\ref{BHH}) \cite{A:Fisher,A:Scalettar,A:Amico,A:Amico2} was introduced 
as model for superconducting films, granular superconductors, 
short-length superconductors, and arrays of Josephson junctions 
\cite{A:Batrouni,A:Fisher2}. 
More recently, some authors \cite{A:Jaksch,A:Stringari,A:IJMP} suggested to
describe the dynamics of ultracold dilute gas of bosonic atoms 
trapped in an optical lattices by means of a BHM. Experimental
results have shown that the essential physics of arrays of coupled 
BECs is captured by  BHM \cite{A:GreinSC,A:Orzel}.
At zero temperature, the ground state of the homogeneous version 
($\lambda_{ j}=const$) of system described by Hamiltonian (\ref{BHH}) 
undergoes a quantum phase transition from the superfluid (SF) phase 
to the Mott insulator (MI) one \cite{A:Fisher,A:Scalettar,A:Amico,A:Amico2,A:Batrouni,A:Fisher2}. 
For values of $T/U$ strong enough the ground-state 
of (\ref{BHH}) is a SF and it is well described by a wave function 
exhibiting a site independent phase \cite{A:Band,A:TrombL2,A:Menotti2}. 
When the lattice potential depth $\omega$ is increased 
and, correspondingly, $T/U$ is decreased the system ground-state  
can manifest two behaviours. If the total number of atoms
$N$ is commensurate with the site number $M$ the system ground-state is 
a MI with vanishing global compressibility, otherwise it is a SF. 
In the case of inhomogeneous potentials, as that resulting from the
confining trap in system (\ref{BHH}), there exist Mott insulating
regions for $T/U$ below a threshold even without commensurate
filling \cite{A:Polkov,CM:Batrouni}.

\section{Superfluid initial state.}

We consider a set of initial conditions in which the Hamiltonian
parameters entail a superfluid ground-state. These are achieved
in the limit where $T/(N U) >> 1$. For $N$ large and $T/(N U)$ over a
suitable threshold, it is widely accepted that the low-temperature 
dynamics of (\ref{BHH}) can be described by a discrete version of GPE
\cite{A:Polkov}. 
This semiclassical limit can be accomplished recalling that 
when the tunneling term dominates the on-site repulsion one 
(namely $T/(UN)>>1$), the Glauber coherent states give rise to 
effective solutions for the quantum problem entailed by Hamiltonian 
(\ref{BHH}) \cite{A:Amico,A:Amico2,A:IJMP}.
Thus, a reasonable solution for the BHM ground-state in the regime of
interest, can be obtained within a {\it coherent state variational picture}
based on applying a time-dependent variational 
principle on coherent-state trial state. 
By means of this procedure (for details see~\cite{A:Zhang,A:Amico,A:Amico2}), the
quantum dynamics generated by Hamiltonian (\ref{BHH}) can be reformulated
in terms of a classical dynamics generated by an effective Hamiltonian 
${\cal H}$. Hence, following this procedure, we assume the system dynamics 
to be described by the trial state $ | \Psi \rangle ={\rm exp}(iS/\hbar) 
|Z\rangle$, where $|Z \rangle := \Pi_i |z_i \rangle$ (see \cite{A:Amico,A:Amico2}) 
is written in terms of Glauber coherent states as
\be
|z_i \rangle : = {\rm e}^{- \frac{1}{2}|z_i |^2}
\sum^{\infty}_{n=0} \frac{z_i^n}{n!} \,
(a^{\dagger}_i )^n |0 \rangle  
\label{CS}
\ee
(recall that $a_i \, |0 \rangle = 0$, and that their defining equation is
$ a_i | z_i \rangle = z_i| z_i  \rangle$ with $z_i \in {\bf C}$).
The effective equations of motion are achieved by a variational principle 
from the effective action $S=\int dt [i \Sigma_j (\dot{z}_j z^*_j -
\dot{z^*}_j z_j)/2 -{\cal H}]$, 
associated to the classical Hamiltonian ${\cal H}( Z, Z^*):= \langle Z| H |Z 
\rangle$, through a variation respect to $z_j$ and $z^*_\ell$.
Hence the time-dependent trial-state parameters $z_j=
\langle \Psi| a_j |\Psi \rangle$
represent the 
classical canonical variables of the effective Hamiltonian dynamics and
satisfy to the Poisson brackets $\{z^*_j,z_\ell  \} = i \delta_{j \ell}
/\hbar$. After some algebra we obtain the classical Hamiltonian
\be
{\cal H} = \sum_{j} \Bigl [ U |z_j|^4  +
\lambda_j |z_j|^2 - \frac{_{T}}{^2}
\left (z^{*}_j z_{j+1} + 
\ {\rm c.c.} \ 
\right ) \Bigr ] \,  ,
\label{HS}
\ee
where $j$ runs on the chain sites: $j\in I_{_M}$ with
$I_{_M}= \{ 0,\pm 1,\ldots , (M-1)/2 \}$ or
$I_{_M}= \{ \pm 1/2,\pm 3/2,\ldots , M/2 \}$ when $M$ is odd or even, 
respectively. The related equations of motion are the following
\bq
 i\hbar {\dot z}_j &=& (2U|z_j|^2 +\lambda_j ) z_j
-\displaystyle \frac{_{T}}{^2} (z_{j-1}+z_{j+1}) \; ,
\label{CEM}
\eq
with $j\in I_{_M}$, and with the complex conjugate equations.

The ground state of Hamiltonian (\ref{HS}) is determined by 
studying the variation of ${\cal H}-\chi {\cal N}$, where the Lagrange 
multiplier $\chi$  has been introduced to explicitly incorporate the 
conserved quantity ${\cal N}=\Sigma_j |z_j|^2$. 
In the limit where the on-site chemical potential $\lambda_j$ is
slowly varying with the site index $j$, namely $|\lambda_j -
 \lambda_{j+1}|/U << 1$, we have $z_{j-1} + z_{j+1} \approx 2 z_j$.
An approximate solution for the SF configuration is thus given by
\be
M^\prime \chi = 2UN + M^\prime \bar{\lambda} - M^\prime T \, , \quad
z_j = \sqrt{\frac{N}{M^\prime}-
\frac{(\lambda_j - \bar{\lambda})}{2 U}}\  e^{i \phi} \, ,
\ee
where $\bar{\lambda}=\Sigma_{_{|j|\in {I_{_{M^\prime}}}}} \lambda_j/ M^\prime$
and $M^\prime=min(M,q)$ is determined by finding out the maximum integer $q$ 
such that $2UN + q (\bar{\lambda} - \lambda_q ) \geq 0$. 
This solution represents the discrete version for the Thomas-Fermi 
approximation \cite{A:DGPS}.
The corresponding energy is
\be
E_{gs} = N /\sigma [ 1 - M^\prime \tau - \sigma (\bar{\lambda} + 
\sigma/4 \bar{(\delta \lambda)^2} ) ]  \, ,
\label{EGSf}
\ee
in which we have set $M^\prime \bar{(\delta \lambda)^2}:=
\Sigma_{_{j\in I_{_{M^\prime}}}} 
( \lambda_j - \bar{\lambda})^2$, $\sigma := M^\prime/(UN)$ and $\tau := 
T/(UN)$.
Concerning the three coupled condensates system within the SF regime, a
thorough study has been made in Ref. \cite{A:PFlet1}.


\section{Superfluid to Mott-insulator transition.}

At the time $t=0$ the optical-lattice potential is suddenly 
increased. This is achieved by varying the potential intensity according to, 
for example, a tilted slope: $\omega (t) = \omega [1+(w-1)t/\tau_b]$, 
where $\tau_b$ is the time scale for the jump and $w$ is the amplification 
factor. 
Consequently, the tunneling amplitude goes to zero $T[\omega(t)]\to 0$ as an 
exponential. In fact, standing the definition $T:=-2 \int d^3{\bf r } 
\bar{u}_{ j} [V - V_{ j\pm { 1}}]u_{{ j} \pm { 1}}$, where $u_{{ j}}$ 
is the harmonic oscillator ground state involved by the quadratic single 
particle potential $V_j$, by direct analytical calculations we get
\be
T[\omega(t)]= \frac{\hbar^2 \omega^2(t)}{4 E_r}
\left[ \frac{\pi^2}{2} - 1 + \frac{2 E_r}{\hbar\omega (t)} - 
e^{- \frac{2 E_r}{\hbar \omega(t)}} \right] 
e^{- \frac{\pi^2\hbar \omega (t)}{8 E_r} } \, .
\label{hopping}
\ee 
Also $U$ and $\lambda_j$ are modified when changing the optical
potential amplitude, but their dependence on $\omega (t)$ is much less 
dramatic. In fact we have $U[\omega(t)] = U [\omega (t)/ \omega]^{1/2}$ and 
$\lambda_j [\omega(t)] = \lambda_j + \hbar [\omega (t) - \omega]/2 $.
While the potential depth is changed, $0< t < \tau_b$, we suppose
the state of the system to be not modified. 

To apply the sudden approximation,
the time scale $\tau_b$ characterising the potential-depth jump 
must be fast compared with the tunneling time between neighbouring wells, 
but slow enough to prevent the condensate excitations in each well, 
namely $ 2 \pi/\omega << \tau_b << \hbar/T $
(where $T$ is the larger hopping amplitude).
In fact, doing so the system persists in the lowest band and the effects 
due to the hopping term result negligible. Thus, a general initial state 
$\sum_{\{\bf n \}} c_{\bf n} | {\bf n}\rangle$, will have the straightforward to
compute time evolution
$\sum_{\{ \bf n \} } \exp\{-i/\hbar \sum_j [U n^2_j +\lambda_j n_j] \} 
c_{\bf n} |{\bf n}\rangle$.
Concerning the sudden approximation, we recall that it is usually used
for calculating transition probabilities in the case when the Hamiltonian 
changes rapidly within a short time interval 
(presently this is identified by $t=0$ and $t=\tau_b> 0$). 
One simply assumes the reaction of the initial state to the quick Hamiltonian 
change to be negligible. 
So one can approximate the transition amplitude by 
assuming: $\langle out |U(\tau_b,0) |in \rangle \approx \langle out 
|in \rangle$. Such an argument works only if impulsive forces are absent, 
which, otherwise, could generate finite states change even if applied for 
infinitesimally long time. In our system such kind of forces are absent. 
In fact, in the time interval $(0,\tau_b)$, where the Hamiltonian parameters 
are time dependent and the jump of the optical-potential depth is driven by 
$w$, by means of the coherent state representation of the path integral, 
we have \cite{B:Inomata}
\be
\langle z| U(\tau_b,0) |z \rangle =
\int {\cal D} [z] 
\exp \left [ \frac{i}{\hbar} \int_0^{\tau_b} \, dt\,  L(t)  \right ]
\, .
\ee
where 
$$
L(t) = \left\{ \frac{i}{2}\sum_k
[ z^*_k(t) \dot{z}_k(t) - \dot{z}^*_k(t) z_k(t)  ] 
- {\cal H}[ z(t), z^*(t),t] \right\} 
$$ 
is the Lagrangian of the effective path-integral action.
Since $L(t)$ can be shown to have no singular behaviours as $\tau_b \to 0$,
namely  $\lim_{\tau_b \to 0^+} \int^{\tau_b}_0 dt L(t) = 0$,
the above formula implies that the dynamical 
evolution of the initial state is driven by $\tau_b$. So the shorter $\tau_b$ 
implies the smaller changes of $U(\tau_b,0) |z \rangle $.

In the Mott regime ($T/U  << 1$), namely after the potential jump 
($t>\tau_b$, we will assume $\tau_b = 0$), the classical description of 
dynamics is no longer valid and the Schr\"odinger equation is necessary
to describe the time evolution. The latter is generated by Hamiltonian 
(\ref{BHH}) where $T=0$, $U=U(w\omega) =:\tilde U$ and $\lambda_j = 
\lambda_j(w\omega) =: \tilde  \lambda_j$ are assumed.
As assumed above, the system initial state is described by 
the coherent state (\ref{CS}) whose the quantum evolution is given by
\be
|(t)\rangle = e^{ - \frac{i}{\hbar} H t} \prod_{i\in I_{_M}} |z_i \rangle
= \prod_{i\in I_{_M}} e^{ - \frac{|z_i|^2}{2}} \sum_{n_i=0}^{+\infty}
\frac{[z_i \ \nu_i(t)]^{n_i}}{\sqrt{n_i !}} e^{-i n^2_i u(t)}
| n_i \rangle~~~~ \, ,
\label{CST}
\ee
where we have set $\nu_i(t):= \exp [i/\hbar (\tilde U+\tilde \lambda_i) t]$ 
and $u(t) =  \tilde U t/ \hbar$. The quantum time evolution does not preserve 
the coherent state structure of state (\ref{CS}), in fact, the term 
$\exp [ -i n^2_i u(t)]$ in 
(\ref{CST}) breaks the coherent state form. 
By direct calculations, it is easy to show that the quantum 
evolution in the Mott-regime time interval entails the following 
expectation values
$$
\langle(t) |a^+_{j} a_{j}|(t)\rangle = |z_j |^2 \, , 
$$
\be
\!\!\!\!\!\!\!\!\!\!\!\!\!\!\!\!\!\!\!\!\!\!\!\!\!\!\!\!\!\!\!\!\!\!\!
z_j(t):=\langle(t) | a_{j}|(t)\rangle = z_j
\exp { \left\{ \frac{i}{\hbar} \tilde \lambda_j t  - 
i|z_j|^2 \sin^2 [u(t)] \right\} }
\exp \left\{-2 |z_j|^2 \sin^2 [u(t)]\right\} \, , 
\label{ipsi}
\ee
$$
\langle(t) |a^+_{j+1} a_{j}|(t)\rangle  = z^*_{j+1} z_j 
\exp \left\{-2 (|z_j|^2 + |z_{j+1}|^2)\sin^2 [u(t)]\right\} \times
$$
\be
\quad \quad \times
\exp \left\{ \frac{i}{\hbar} (\tilde \lambda_j - \tilde \lambda_{j+1})t -
 i (|z_j|^2 - |z_{j+1}|^2)\sin^2 [u(t)] 
 \right\}  \, .
\label{ics}
\ee
Such equations display that, during the quantum time evolution,
the wells population does not change, whereas the site wave-functions
$z_j(t)$ are dynamically active and driven by more then one characteristic 
times.
The modulus of $z_j(t)$ is a periodic function of~$t$ 
$\vert z_j(t) \vert = \vert z_j \vert \exp \{-2 |z_j|^2 \sin^2 
(\tilde U t/\hbar)\}$ with period $T_m = \pi \hbar/ \tilde U$.
\begin{figure}[h]
\begin{center}
\includegraphics[width=7.5cm]{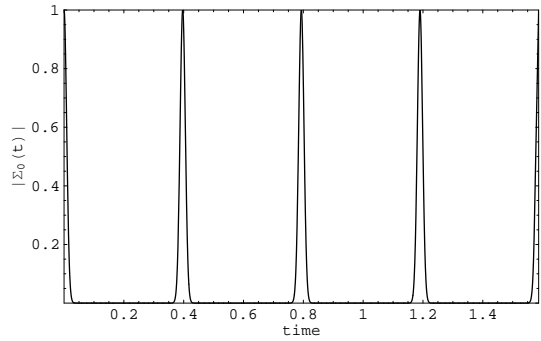}
\caption{\label{Fig1} This figure shows the time-evolution of the modulus of
the central-site wave-function $\Sigma_0(t) = z_0(t)/z_0$ as given by Eq. 
(\ref{ipsi}). It displays a periodic behavior whose period is $T_m = \pi \hbar/ 
\tilde U \approx 0.396$ sec. 
Notice the ``almost impulsive'' periodic behavior  of $\Sigma_0 (t)$ which
is mostly almost vanishing. 
}
\end{center}
\end{figure}
The phase of the site wave-functions $\varphi_j := \arg 
[z_j(t)/\vert z_j(t) \vert] = \tilde \lambda_j t/{\hbar}   - 
|z_j|^2 \sin^2 [u(t)]$ is driven by $T_m$, the period of $\sin^2 [u(t)]$,
and by the $T_{\tilde \lambda_j} = 2 \pi \hbar /\tilde \lambda_j$ the site 
dependent periods involved by the external potential $\tilde \lambda_j$. 
Furthermore, in each site $j$ where $|z_j|^2$ exceeds the value $2 \pi$ 
the condition $|z_j|^2 \sin^2 [u(t+T_{z_j})] = |z_j|^2 \sin^2 [u(t)] + 
2 \pi$ implies a further characteristic time $T_{z_j}$ for $j\in I_M$.
\begin{figure}[h]
\begin{center}
\begin{tabular}{cc}
\includegraphics[width=7.5cm]{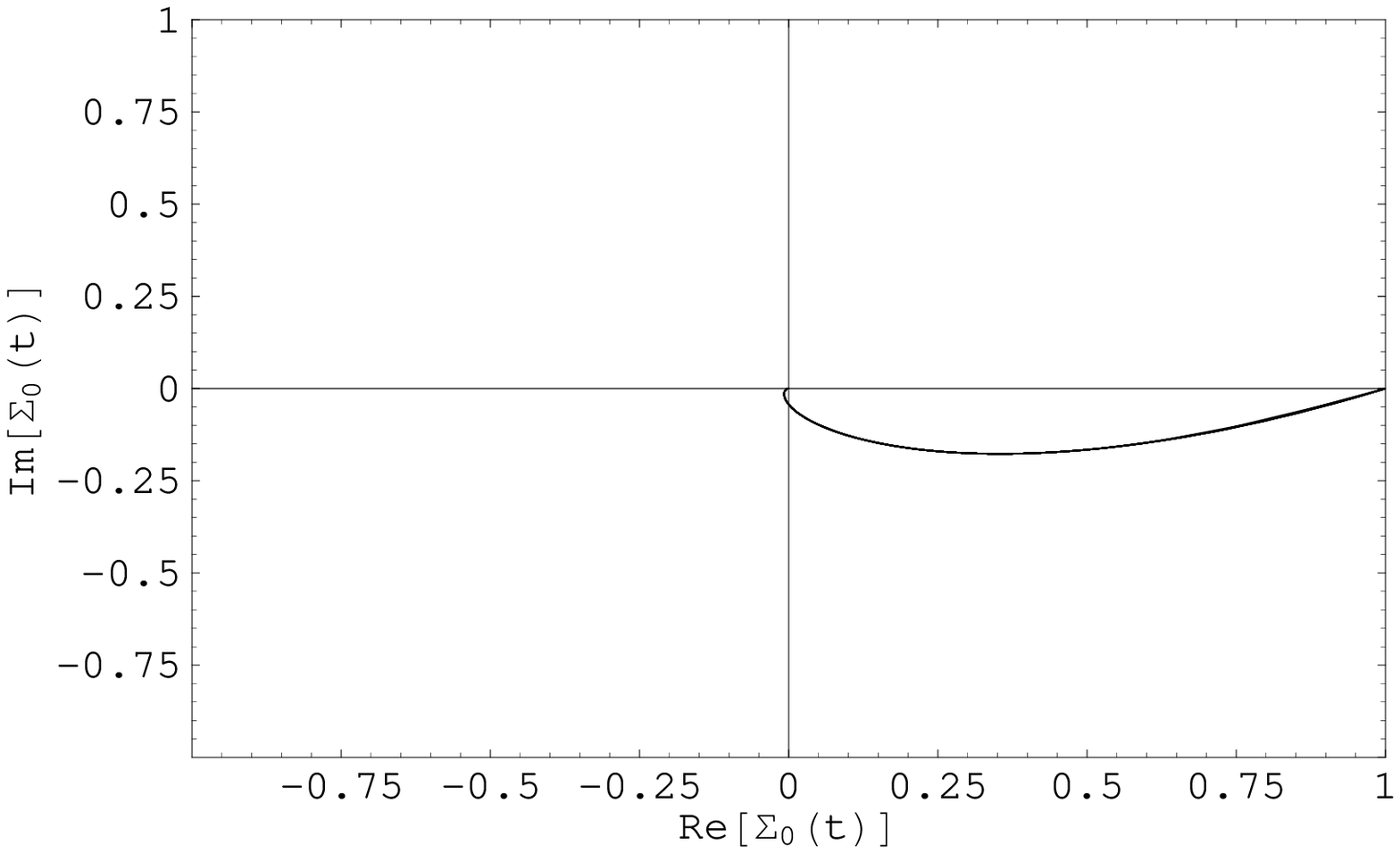}&
\includegraphics[width=7.5cm]{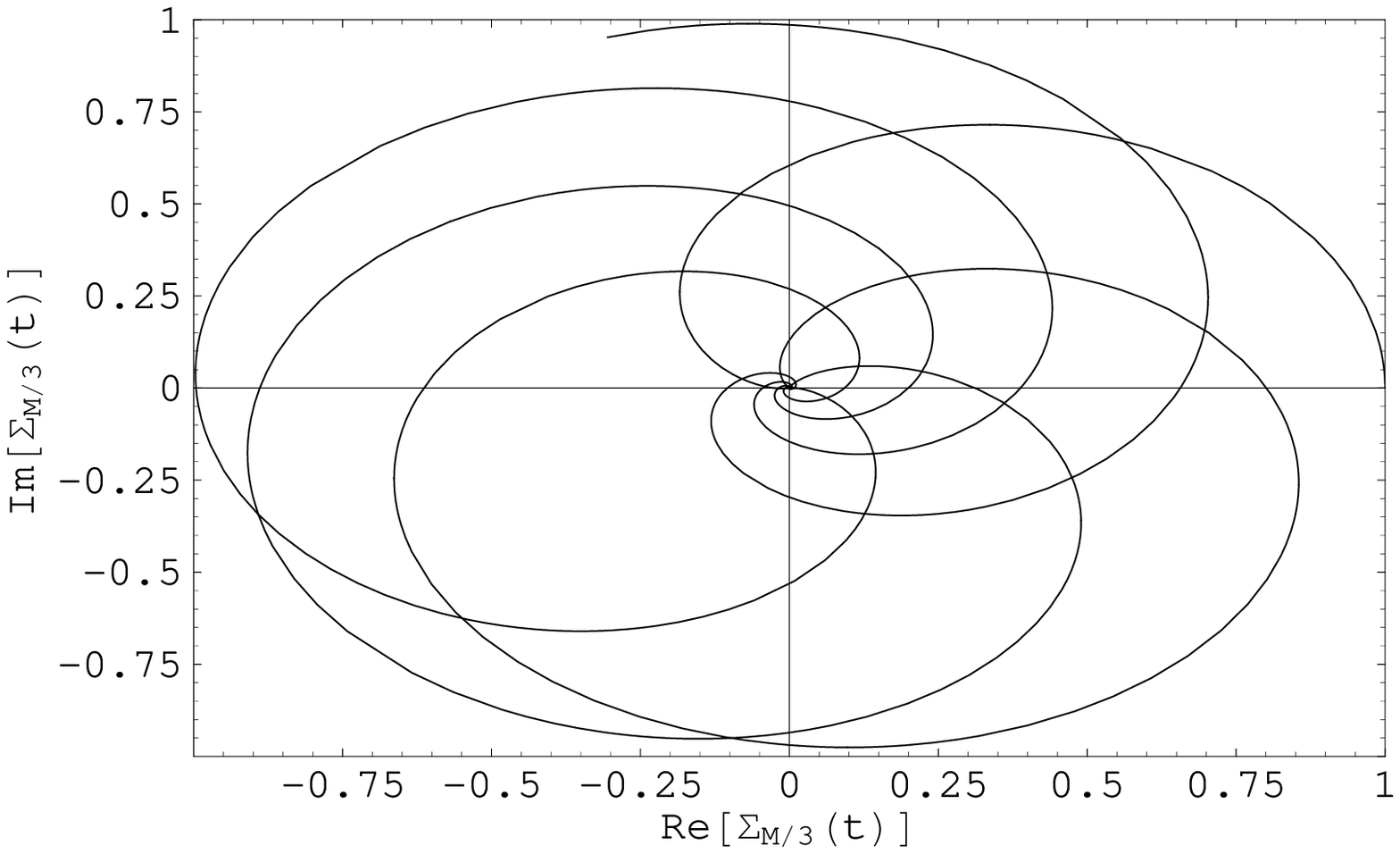}
\end{tabular}
\caption{\label{Figs23} Trajectories of the site wave-functions
$\Sigma_j (t)= z_j(t)/z_j$ in the complex plane (abscissae and ordinates
refer to real and imaginary part, respectively) for $0< t< 4T_m$. 
Left and right figures refer to site  $j=0$ (the central one) and site 
$j= M/3$, respectively.  Notice that, unlike $\Sigma_0 (t)$, $\Sigma_{M/3}(t)$ 
is not periodic.
}
\end{center}
\end{figure}

Also, the site-dependent external potentials $\tilde \lambda_j$
induce a dephasing between the $z_j(t)$ and $z_{j+1}(t)$ representing 
the condensate states (namely the site wave function) at sites $j$ and $j+1$.
In fact, it results $\varphi_j - \varphi_{j+1} = - 
(\pi \hbar \Omega_1/2)^2 (2j+1)t/(\hbar E_r) - (\vert z_j \vert^2 -
\vert z_{j+1} \vert^2) \sin^2 [u(t)]$ that shows as the dephasing 
increases along the lattice.
\begin{figure}[h]
\begin{center}
\includegraphics[width=7.5cm]{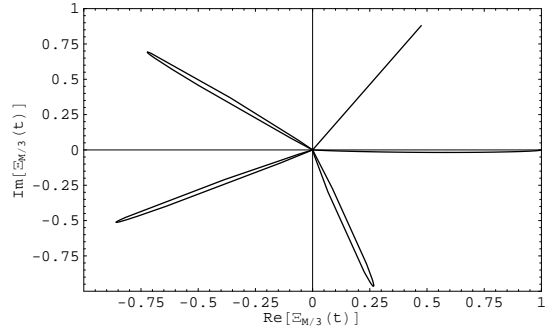}
\caption{\label{Fig4} The evolution of the spatial correlation of
adjacent sites is represented by the parametric plot of $\Xi_j (t) = 
\langle(t) |a^+_{j+1} a_{j}|(t)\rangle  / z^*_{j+1} z_j$, with $j=M/3$ 
and $0<t< 4 T_m$. A strong dephasing between the sites under concern is 
evident.
}
\end{center}
\end{figure}
The difference between the phases of the condensates 
announces that the system is no more in the ground-state
\cite{A:Band,A:TrombL2,A:Menotti2} and give rise to excited configurations.

Figures \ref{Fig1}-\ref{Fig4} have been achieved
by considering e realistic experimental configuration with $10^4$ $^{85}$Rb 
atoms, a harmonic trapping potential frequency $\Omega_j \approx 50$ Hz 
(j=1,2,3), and a laser mode $k \approx 10^7$ m. The optical potential 
amplitude, initially of the order $E_r$, is suddenly increased to $30 E_r$.


\section{Mott-insulator to superfluid transition.}

Times $t > t^\prime$ correspond to the third stage of the dynamics.
The optical-potential depth is quickly decreased (in a time of order 
$\tau_b \approx 0$) to the original value $\omega$ thus restoring
the regime with $T/(N U) >> 1$. In this case an approximate 
description of the system dynamics within the semiclassical variational 
picture applied in the superfluid regime should be again applicable. 
The initial conditions for the third stage of the system evolution can
be easily shown to be represented by a superposition of Glauber's
coherent states. 
In fact, the initial state is obtained from Eq. (\ref{CST}) 
by setting $t=t^\prime$
\[
|(t^\prime)\rangle = 
\prod_{j\in I_{_M}} {\cal E}_j \sum_{n_j=0}^{+\infty}
\frac{[z^\prime_j]^{n_j}}{\sqrt{n_j !}} e^{-i n^2_j u^\prime}
| n_j \rangle~~~~ \, .
\]
Here ${\cal E}_j = \exp \{ - |z_j|^2/2 \}$, 
$z^\prime_j = z_j \nu_j(t^\prime)$, and $u^\prime = u(t^\prime)$.
By using the identity 
\[
\lim_{\epsilon \to 0^+}
\int^\infty_{-\infty} d x \exp 
[-(p + \epsilon) x^2 - in x] = \exp [- n^2/(4 p)] \sqrt{ \pi/p}  \, ,
\]
with $p=-i/(4u^\prime)$, this state can be written in a very suggestive form
as a superposition of coherent states. Direct calculations give
\be
\!\!\!\!\!\!\!\!\!\!\!\!\!\!\!\!\!\!\!\!\!\!\!\!\!\!\!\!\!\!\!\!
|(t^\prime)\rangle = 
\prod_{j\in I_{_M}}  
\int^\infty_{-\infty}  \frac{d x_j}{2 \sqrt{\pi u^\prime}}
e^{-i \pi /4 } 
e^{i x^2_j / (4 u^\prime)} |  z^\prime_j e^{- i x_j }\rangle 
= 
\prod_{j\in I_{_M}}  
\int^\infty_{-\infty}  d x_j
K(x_j, u^\prime) |  z^\prime_j e^{- i x_j }\rangle 
\, ,
\label{sup}
\ee
where the state labeled by $z^{\prime\prime}_j=z^\prime_j e^{- i x_j }$ 
is the normalized coherent state given in Eq. (\ref{CS}) with 
$z_j = z^{\prime\prime}_j$. 
If one expresses $|(t^\prime)\rangle$ as $|(t^\prime)\rangle = 
\prod_{j\in I_{_M}} |(t^\prime)_j\rangle $, the new trial state accounting
for the evolution after the second potential-depth change, might be
expressed as
$$
|(\xi_j) \rangle = \prod_{j\in I_{_M}} D(\xi_j)\, |(t^\prime)_j\rangle 
$$ 
with $D(\xi_j) = \exp (\xi_j a^+_j - a_j \xi^*_j)$,
where the time behaviour of new dynamical parameters $\xi_j$, $\xi^*_j $
occurring in the exponential terms (actually these are coherent-state
displacement operators) must be reconstructed by implementing once more
the time-dependent variational procedure. 
Due to the properties characterising 
the the displacement-operators action on a coherent state $| z \rangle$ 
($D(\xi) | z \rangle = \exp [iIm(z^* \xi)] | z+\xi \rangle$) the final
form of the trial state $|(\xi_j) \rangle$ is
$$
|\{ \xi_j(t) \} \rangle= \prod_{j\in I_{_M}}  
\int^\infty_{-\infty}  d x_j
K(x_j, u^\prime) e^{\phi_j } \,| \xi_j (t) + z^\prime_j e^{- i x_j }\rangle 
$$
where the kernel $K(x_j, u^\prime)$ is defined implicitly, 
and $\xi_j (t) = 0$ at $t=0$ and $\phi_j = 
Im [\xi^*_j (t) z^\prime_j e^{- i x_j }]$. 



\section{Final discussion.}

In the present paper we have considered a performable experimental
process exhibiting a strong interplay between classical (SF)
and quantum (MI) regimes. We have described the dynamics of an array of 
BECs when the optical potential depth is quickly varied. The process
we have considered forces the system to go through an intermediate 
quantum regime. As a consequence of this, the system loses its 
semiclassical character assuming the form of an excited state 
that cannot be represented as a simple direct product of coherent states. 
Eqs. (\ref{ics}) show that collapsing/revival
phenomena occur whose characteristic time scales have been recognized. 
Further, the presence of the harmonic external 
potential appears to responsible for a strong site dephasing.
When the optical potential depth is lowered again, the resulting state
has been shown, within the previous section, to be a superposition of coherent 
states represented by the integrals of eq. (\ref{sup}).
In performing the integration on $x_j$, at each site, each coherent state
$|  z^\prime_j e^{- i x_j }\rangle$ contributes with a phase $e^{- i x_j }$.
The latter might have a destructive effect for increasing $u^\prime$
when calculating the expectation value of the physically relevant operators 
of the model. This can be seen by re-expressing the integrals in (\ref{sup}) 
in the form 
\[
\int_{-\infty}^{\infty} 
\frac{d x}{\sqrt{\pi u^\prime}}  e^{i x^2 / (4 u^\prime)} 
e^{ \exp [(- i x ) z^\prime a^\dagger] } \, .
\]
Also, since the term $\exp [i x^2 / (4 u^\prime)]/\sqrt{u^\prime}$
is rapidly oscillating outside the interval $[-\sqrt{\pi u^\prime},
\sqrt{\pi u^\prime}]$, the major contributions is expected to come 
from the integration on this interval. Observing that the corresponding 
coherent-states phases change in $- \sqrt{\pi u^\prime} \leq x 
\leq \sqrt{\pi u^\prime}$,  the simplest way to reduce the 
decoherence effects may be achieved by imposing $u^\prime << 1$, 
that is $t^\prime <<\hbar/\tilde{U}$. In view of their complexity 
this problem and the evaluation of the time behaviour of the state 
$|\{ \xi_j(t) \} \rangle$ emerging from the Mott regime will be
discussed in a separate paper.

\ack

R. F. wishes to thank Prof. Ennio Arimondo's BEC group in Pisa
for comments and suggestions.

\section*{References}

\begin{thebibliography}{10}

\bibitem{A:MHAnderson}
Anderson M~H, Ensher J~R, Matthews M~R, Wiemann C~E and Cornell E~A 1995 {\em
  Science\/} {\bf 269} 198

\bibitem{A:Brandley}
Brandley C, Sackett C~A, Tollet J~J, Anderson R~G~H~B and Kasevich M 1995 {\em
  Phys. Rev. Lett.\/} {\bf 75} 1687

\bibitem{A:BAnderson}
Anderson B and Kasevich M 1998 {\em Science\/} {\bf 282} 1686

\bibitem{A:Marzlin}
Marzlin K and Zhang W 2000 {\em Eur. Phys. J. D\/} {\bf 12} 241

\bibitem{A:Morsch}
Morsch O, Muller J~H, Cristiani M, Ciampini D and Arimondo E 2001 {\em Phys.
  Rev. Lett.\/} {\bf 87} 140402

\bibitem{A:Morsch2}
Morsch O, Muller J~H, Cristiani M, Ciampini D and Arimondo E 2002 {\em Phys.
  Rev. A\/} {\bf 65} 06312

\bibitem{A:Berg}
Berg-S{\o}rensen K and M{\o}olmer K 1998 {\em Phys. Rev. A\/} {\bf 58} 1480

\bibitem{A:GreinSC}
Greiner M, Mandel O, Esslinger T, Hansch T~W and Bloch I 2002 {\em Science\/}
  {\bf 415} 39

\bibitem{A:Orzel}
Orzel C, Tuchman A~K, Fenselau M, Yasuda M and Kasevich M~A 2001 {\em
  Science\/} {\bf 291} 2386

\bibitem{A:Polkov}
Polkovnikov A, Sachdev S and Girvin S 2002 {\em Phys. Rev. A\/} {\bf 66} 053607

\bibitem{A:Celeghini}
Celeghini E and Rasetti M 1998 {\em Phys. Rev. Lett.\/} {\bf 80} 3424

\bibitem{A:Chernyak}
Chernyak V, Choi S and Mukamel S 2003 {\em Phys. Rev. A\/} {\bf 67} 053604

\bibitem{A:Band}
Band Y and Trippenbach M 2002 {\em Phys. Rev. A\/} {\bf 65} 053602

\bibitem{A:TrombL2}
Trombettoni A, Smerzi A and Bishop A~R 2001 {\em Phys. Rev. Lett.\/} {\bf 88}
  173902

\bibitem{A:Menotti2}
Kr{\"{a}}mer C~M~M, Pitaevskii L and Stringari S 2003 {\em Phys. Rev. A\/} {\bf
  67} 053609

\bibitem{A:Milburn}
Milburn G~J, Corney J, Wright E~M and Walls D~F 1997 {\em Phys. Rev. A\/} {\bf
  55} 4318

\bibitem{A:Fisher}
Fisher M, Weichman P~B, Grinstein G and Fisher D~S 1989 {\em Phys. Rev. B\/}
  {\bf 40} 546

\bibitem{A:Scalettar}
Scalettar R, Batrouni G, Kampf A and Zimanyi G 1995 {\em Phys. Rev. B\/} {\bf
  51} 8467

\bibitem{A:Amico}
Amico L and Penna V 1998 {\em Phys. Rev. Lett.\/} {\bf 80} 2189

\bibitem{A:Amico2}
Amico L and Penna V 2000 {\em Phys. Rev. B\/} {\bf 62} 1224

\bibitem{A:Batrouni}
Batrouni G, Scalettar R and Zimanyi G 1990 {\em Phys. Rev. Lett.\/} {\bf 65}
  1765

\bibitem{A:Fisher2}
Fisher M, Grinstein G and Girvin S~M 1990 {\em Phys. Rev. Lett.\/} {\bf 64} 587

\bibitem{A:Jaksch}
Jaksch D, Bruder C, Cirac J, Gardiner C and Zoller P 1998 {\em Phys. Rev.
  Lett.\/} {\bf 81} 3108

\bibitem{A:Stringari}
Stringari S 2001 {\em C. R. Acad. Sci.\/} {\bf 4} 381--397

\bibitem{A:IJMP}
Franzosi R, Penna V and Zecchina R 2000 {\em Int. J. Mod. Phys. B\/} {\bf 14}
  943

\bibitem{CM:Batrouni}
Batrouni G~G, Rousseau V, Scalettar R~T, Rigol M, Muramatsu A, Denteneer P~J~H
  and Troyer M.
\newblock Cond-mat/0203082

\bibitem{A:Zhang}
Zhang W~M, Feng D~H and Gilmore R 1990 {\em Rev. Mod. Phys\/} {\bf 62} 867

\bibitem{A:DGPS}
Dalfovo F, Giorgini S, Pitaevskii L~P and Stringari S 1999 {\em Rev. Mod.
  Phys.\/} {\bf 71} 463

\bibitem{A:PFlet1}
Buonsante P, Franzosi R and Penna V 2003 {\em Phys. Rev. Lett.\/} {\bf 90}
  050404

\bibitem{B:Inomata}
Inomata A, Kuratsuji H and Jerry C 1992 {\em Path Integral and Coherent States
  of SU(2) and SU(1,1)\/} (World Scientific)

\end{thebibliography}

\end{document}